# Solar Corona Thermal Background and Energy Spectrum of the Weak Solar Soft X-ray Bursts

I. K. Mirzoeva

Space Research Institute, Russian Academy of Sciences,

ul. Profsoyuznaya 84/32, Moscow, 117997 Russia

e-mail: colombo2006@mail.ru

**Abstract**

We analyze the time profiles for weak solar soft X-ray bursts over the period 1995–1999 within the framework of the Interball–Tail Probe project. We have revealed a tendency for the peak of the energy spectrum of the time profiles for weak solar soft X-ray bursts to shift and their correlation with the thermal background.

## INTRODUCTION

Studies of low-intensity solar X-ray bursts can provide valuable information about the microstructure of solar active regions. Within the framework of the Interball—Tail Probe project, the solar soft X-ray radiation was measured using the RF-15I-2 X-ray detector (Farnik et al. 1995; Likin et al. 1998). Recall that low-intensity solar events are all events with peak X-ray fluxes $<10^{-6}$ W m$^{-2}$, i.e., class B, A, and 0 events with peak fluxes $<10^{-8}$ W m$^{-2}$ (Mirzoeva and Likin 2002, 2004). Such series of weak X-ray bursts were recorded from September through December 1995 by the RF-15I-2 X-ray detector in the energy range 2–15 keV.

Using data obtained from August 1995 through December 1999, we analyze class A and class 0 bursts. The separation of the energy range of the RF-15I-2 detector into subranges, 2–3, 3–5, 5–8 and 10–15 keV, allowed us to determine the peculiarities of the time profiles for weak bursts in narrower energy bands.

## ANALYSIS OF THE EXPERIMENTAL DATA

For weak solar soft X-ray bursts recorded from August 1995 through December 1999, we found the following peculiarities: in years close to the solar minimum or, more specifically, in 1995 and 1996, most of the class A and 0 bursts were recorded in 2-3 and 3-5 keV energy bands, while the number of such bursts in the 5–8 and 10–15 keV energy bands was insignificant (less than 10% of the total number of bursts). Below, we give several typical examples of the time profiles for weak solar soft X-ray bursts for this period.

Series of class A bursts are seen in the 2–3 and 3–5 keV energy bands from 14 h 46 min December 11 until 11 h 32 min December 12, 1995. The 5–8 and 10–15 keV channels for this period contain only a thermal background (Fig. 1). Series of class 0 bursts and one class A burst were recorded in the same energy bands from 0 h 0 min until 12 h 48 min December 15, 1995. As in the previous case, the 5–8 and 10–15 keV channels for this period contain only a thermal background (Fig. 2). Series of weak class A bursts and series of class B bursts (with X-ray fluxes from $10^{-7}$ to $10^{-6}$ W m$^{-2}$ ) were recorded in the 2–3 and 3–5 keV energy bands from 7 h 01 min May 7 until 0 h 34 min May 8, 1996. The same series of class B bursts were recorded in the 5–8

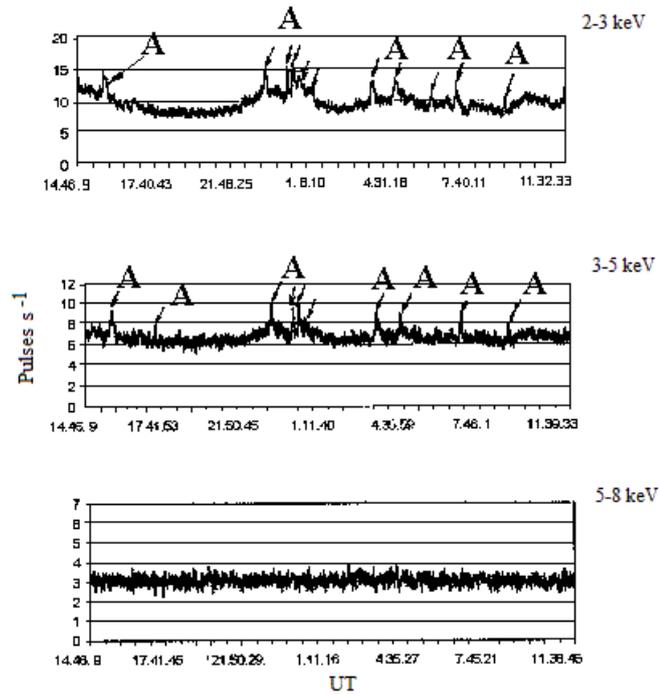

**Fig. 1.** Weak solar soft X-ray bursts recorded in the 2–3 and 3–5 keV energy bands from 14 h 46 min December 11 until 11 h 32 min December 12, 1995. Class A bursts are marked by the arrows.

**Thermal background (pulses s$^{-1}$)**

| Date | | Energy band, keV | | | |
| --- | --- | --- | --- | --- | --- |
| | | 2-3 | 3–5 | 5-8 | 10-15 |
| | 1995 | 7 – 30 | 6 – 10 | 3 | 15 - 27 |
| | 1996 | 7 – 20 | 6 – 10 | 3 | 12 - 18 |
| | 1997 | 6 – 13 | 12 – 70 | 5 – 160 | 9 – 30 |
| January-June | 1998-1998 | 4 – 5 | 6 – 20 | 12 – 90 | 9 - 10 |
| June-December | 1998-1998 | 3 – 4 | 5 - 6 | 5 – 20 | 10 - 50 |
| | 1999 | 3 – 6 | 4 – 9 | 3 – 10 | 9 – 60 |

and 10–15 keV bands, while virtually no class A bursts were observed in these energy bands (Fig. 3).

A shift in the maximum of the number of weak X-ray bursts from the 2–5 to the 3–8 keV band was observed in similar measurements performed in 1997. Meanwhile, very few or no X-ray bursts were observed at energies >8 keV. For instance, weak class A bursts were recorded in the 3–5 and 5–8 keV energy bands from 20 h 01 min June 5 until 12 h 02 min June 6, 1997. The 2–3 and 10–15 keV channels for the same period contain only a thermal background (Fig. 4).

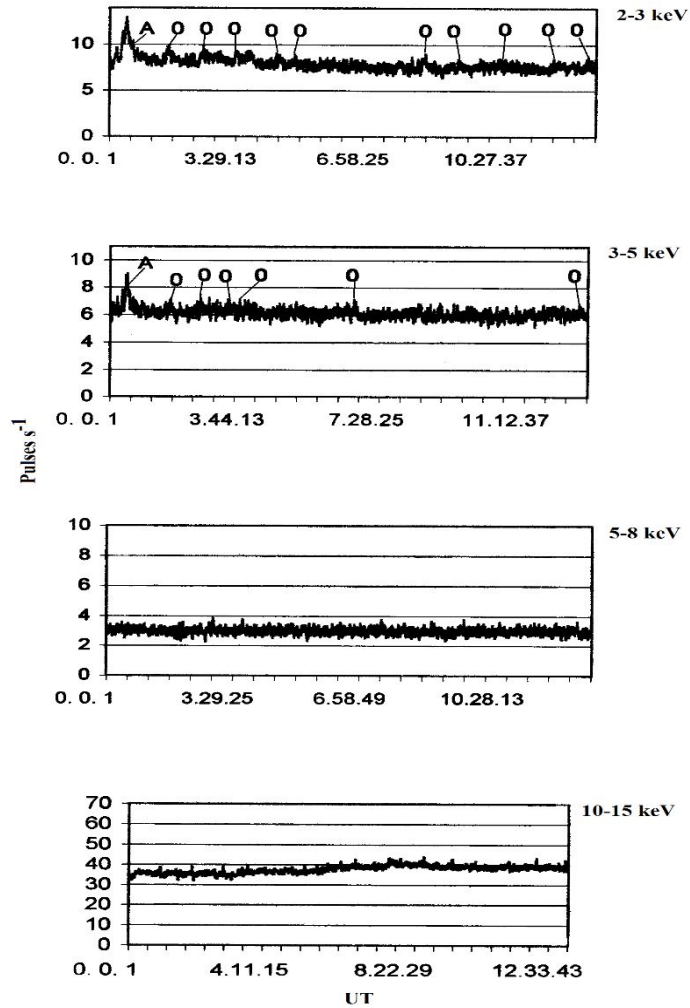

**Fig. 2.** Weak solar soft X-ray bursts recorded in the 2–3 and 3–5 keV energy bands from 0 h 0 min until 12 h 48 min December 15, 1995. Class 0 bursts and a class A burst are marked by the arrows.

We may assume that the peak of the distribution of weak X-ray solar bursts would shift further toward the energies 10–15 keV. Indeed, this assumption was confirmed by the observation of weak bursts in 1998 and 1999. Beginning from the second half of 1998, the picture gradually changed: the number of weak bursts increased in the 5–8 and 10–15 keV energy bands (especially in the 10–15 keV band); i.e., bursts were observed more commonly at relatively high energies. For instance, from 6 h 36 min until 21 h 24 min July 17, 1998, class A bursts were recorded more commonly in the 5–8 and 10–15 keV bands (Fig. 5).

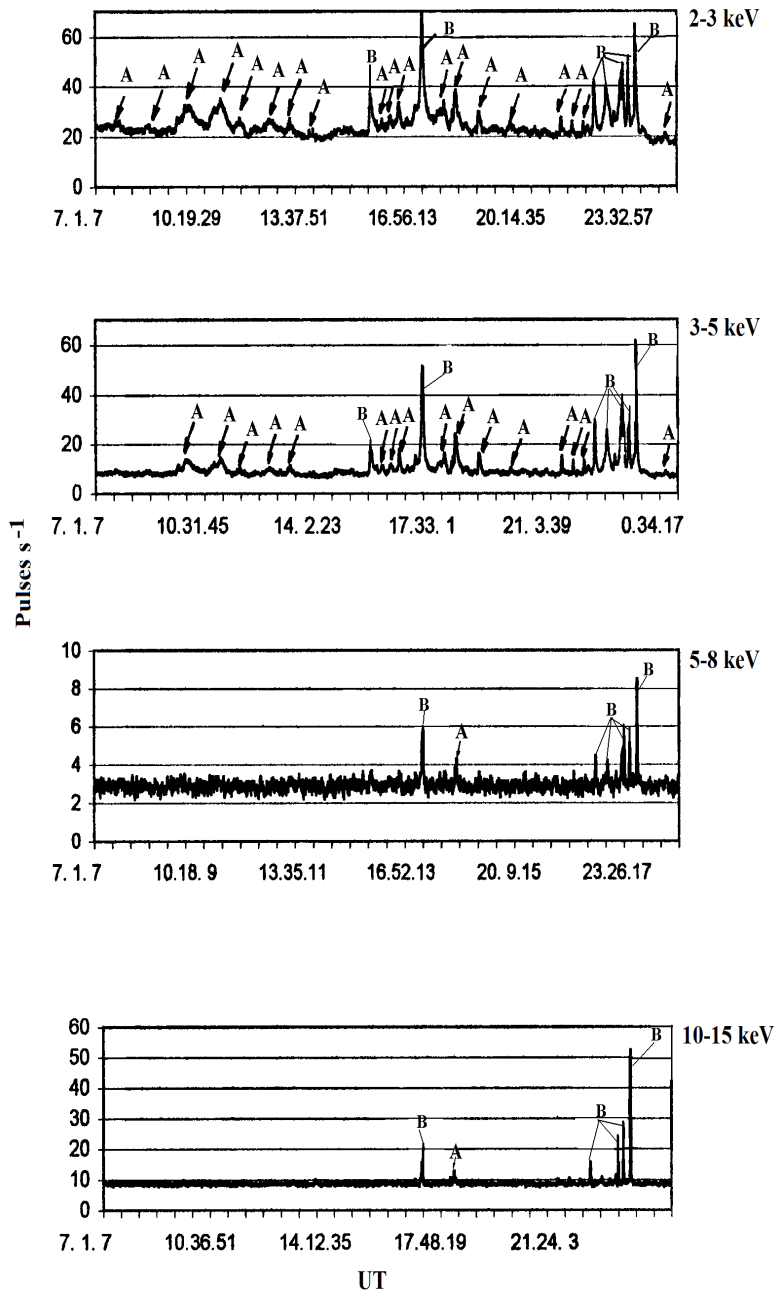

**Fig. 3.** Weak solar soft X-ray bursts recorded in the 2–3 and 3–5 keV energy bands from 7 h 01 min May 7 until 0 h 34 min May 8, 1996. Class A and B bursts are marked by the arrows.

The year 1999 was one close to the maximum of the current solar cycle. In this period, much more strong class C and M solar soft X-ray bursts were observed in all energy bands. This severely hampered the detection of low-intensity solar events. However, on several days when there were few strong bursts, we were able to record the periods suitable for an analysis when weak bursts were detected only in the 10– 15 keV energy band, while there were no weak bursts in the remaining energy bands. Thus, for example, more class A and B bursts were recorded in the 10–15 keV band from 23 h 40 min November 24 until 3 h 53 min November 25, 1999. In the 2–8 keV band, we recorded appreciably fewer such bursts (Fig. 6). This trend persisted throughout 1999.

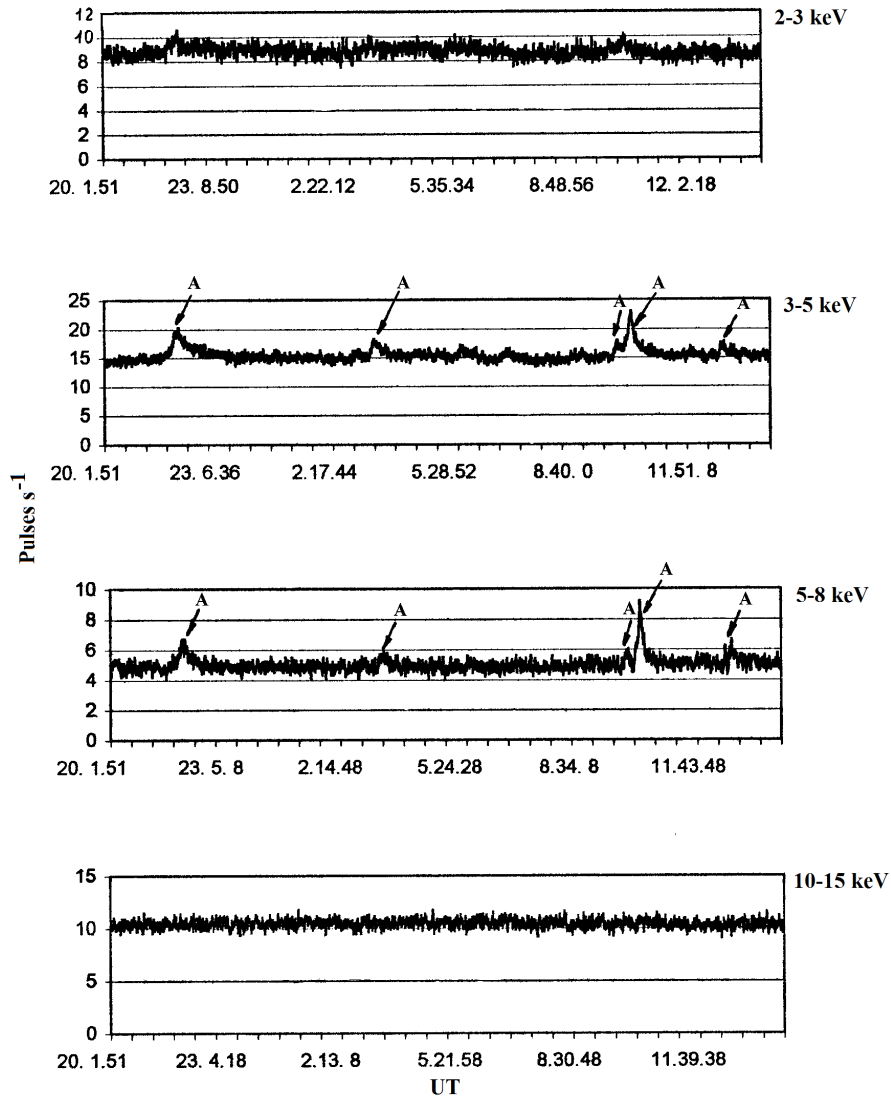

**Fig. 4.** Weak solar soft X-ray bursts recorded in the 3–5 and 5–8 keV energy bands from 20 h 01 min June 5 until 12 h 02 min June 6, 1997. Class A bursts are marked by the arrows.

Figure 7 shows the overall pattern of displacement of the energy spectrum of the time profiles for weak solar soft X-ray bursts in the period 1995–1999. The burst detection energy band in keV and the years of observation are along the vertical and horizontal axes, respectively.

Weak bursts have a significant effect on the background X-ray radiation that is mainly thermal in origin. Thus, during each year of the current solar cycle, the thermal background level was stabilized in the energy bands where few or no weak bursts were observed. At the same time, we recorded a larger spread in thermal background values in the energy bands where more weak bursts were detected. The table gives the spread in mean thermal background values in various energy bands for the period from 1995 until 1999.

In 1995 and 1996, weak bursts were recorded more commonly in the 2–3 and 3–5 keV energy bands. The minimum and maximum thermal background values in these energy bands differed by factors of 4.2 and 1.6, respectively. At the same time, the thermal background in the 5–8 keV band was almost constant, while the spread in thermal background values in the 10–15 keV band decreased from a factor of 1.8 to a factor of 1.5 from 1995 until 1996.

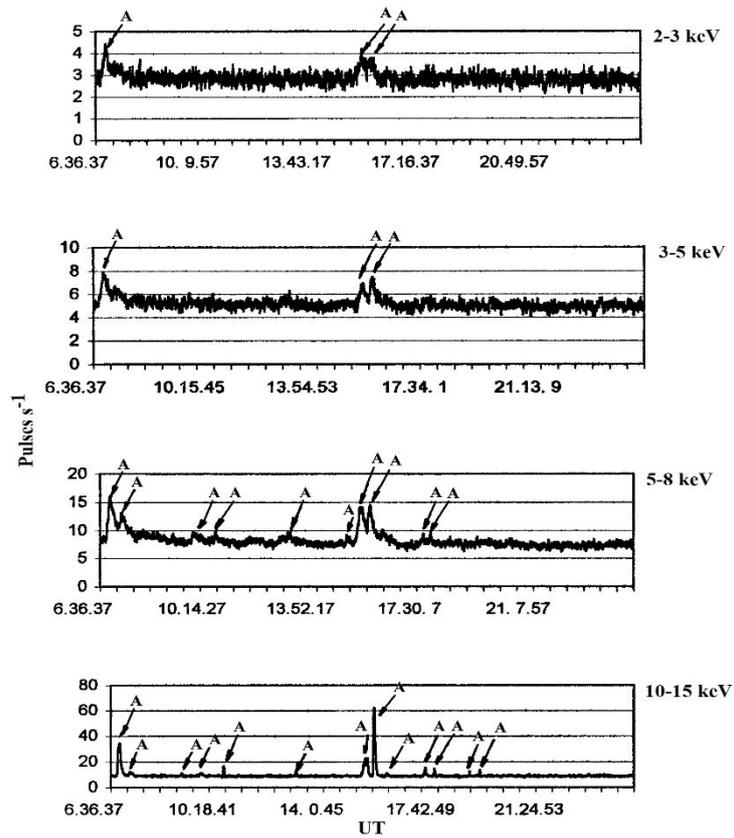

**Fig. 5.** Weak solar soft X-ray bursts recorded in the 5–8 and 10–15 keV energy bands from 6 h 36 min until 21 h 24 min July 17, 1998. Class A bursts are marked by the arrows.

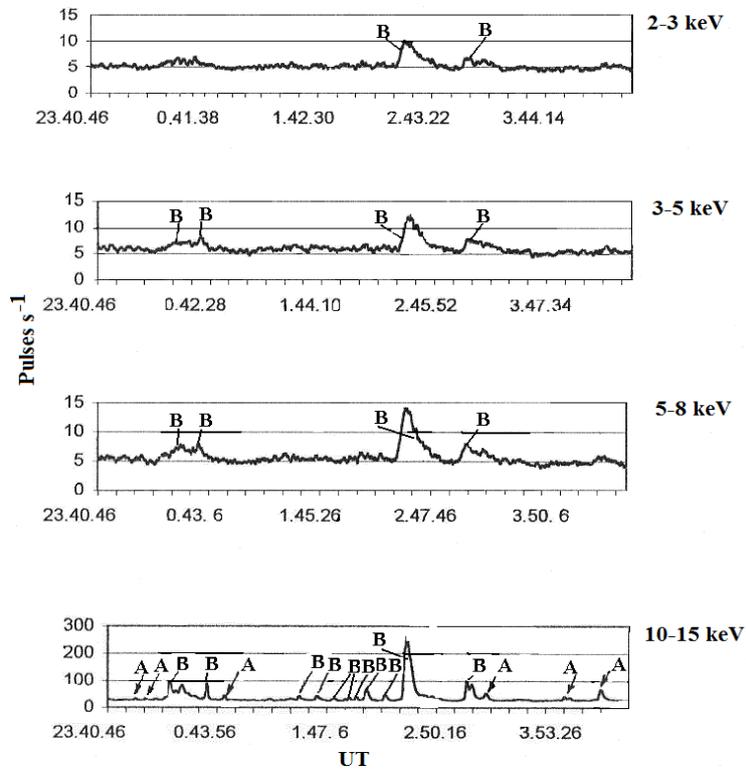

**Fig. 6.** Weak solar soft X-ray bursts recorded in the 10–15 keV energy band from 23 h 40 min November 24 until 3 h 53 min November 25, 1999. Class A and B bursts are marked by the arrows.

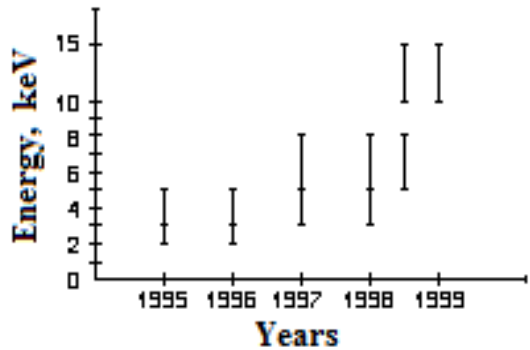

**Fig. 7.** Overall pattern of displacement of the energy spectrum of the time profiles for weak solar soft X-ray bursts in the period 1995–1999.

In 1997, more weak bursts were recorded in the 3–5 and 5–8 keV energy bands. The minimum and maximum thermal background values in these energy bands differed by factors of 6 and 32, respectively, while in the 2–3 and 10–15 keV energy bands where few weak bursts were observed in 1997, the thermal background values differed by factors of only 2 and 3, respectively. In the period January–June 1998, weak bursts were recorded more commonly in the 3–5 and 5–8 keV energy bands, as before. The spread in thermal background values in this part of the energy spectrum decreased slightly: the corresponding values differed by factors of 3 and 8, but this is still larger than the background spread in the 2–3 and 10–15 keV energy bands where there were few weak bursts: here, the values differed by a factor of 1.1. In the period June–December 1998, the energy spectrum of the detected weak bursts shifted toward the energies 5–8 and 10–15 keV. Accordingly, in these energy bands, the spread in thermal background values increased by factors of 4 and 5. In contrast, in the 2–3 and 3–5 keV bands, the background was stabilized, and its spread was much smaller (factors of 1.3 and 1.2, respectively). In 1999, weak bursts were recorded more commonly in the 10–15 keV band. Accordingly, in this energy band, the spread in thermal background values is largest for 1999, a factor of 6.7. In contrast, in the energy bands with a smaller number of weak bursts (2–3, 3–5, and 5–8 keV), the spread in thermal background values was smaller (factors of 2, 2.3, and 3.3, respectively).

**CONCLUSIONS**

1. The number of weak solar soft X-ray bursts in various energy bands from 2–3 to 10–15 keV depended on the observing period in the solar cycle.

2. Depending on the period in the solar cycle, weak bursts were recorded more commonly in the following energy bands:

1995—2–3, 3–5 keV;

1996—2–3, 3–5 keV;

1997—3–5, 5–8 keV;

January–June 1998—3–5, 5–8 keV;

June–December 1998—5–8, 10–15 keV;

1999—10–15 keV.

3. The observations of the time profiles for weak solar soft X-ray bursts, carried out from 1995 until 1999, lead us to the following conclusions about the peculiarities of the physical processes in solar active regions in relatively narrow energy bands in separate periods of the solar cycle:

3.1. The mean energy of weak bursts increases with growing solar activity.

3.2. We found a correlation between the number of weak bursts and the spread in thermal background values: a larger spread in thermal background values is observed in the periods and energy bands where class A and 0 bursts were recorded more commonly.

4. The correlation between the intensity of the background X-ray radiation and weak X-ray bursts can probably be explained by the absence of a sharp boundary between the thermal background and the X-ray burst proper generated by the particle acceleration during a flare in an active region. An X-ray flare can be imagined as a superposition of the thermal and nonthermal X-ray components (Krucker et al. 2002) or as a superposition of a number of elementary energy release events (Pisarenko and Likin 1995), each of which can be either thermal or nonthermal in origin. In any case, we are dealing with the transfer of energy in the flare–background system.

**REFERENCES**

1. F. Farnik, J. Sylwester, and O. Likin, Interball Mis- sion and Payload (Russia, 1995), p. 256. 2. S.Krucker,S.Christe,R.P.Lin,et al.,SolarPhys. 210, 445 (2002).

3. O. B. Likin, N. F. Pisarenko, F. Farnik, et al., Kosm. Issled. 36, 305 (1998).

4. I. K. Mirzoeva and O. B. Likin, Preprint No. 2046, IKI RAN (Inst. Cosm. Res., Russ. Acad. Sci., Moscow, 2002).

5. I. K.Mirzoeva and O. B. Likin, Pis'ma Astron. Zh. 30, 216 (2004).

6. N. F. Pisarenko and O. B. Likin, Izv. Ross. Akad. Nauk 59, 37 (1995).